\documentstyle{article}
\newcommand{\bra}[1]{\left \langle #1 \right \vert }
\newcommand{\ket}[1]{\left \vert #1 \right \rangle}

\begin{document}
%\fleqn
%\draft
\title{Three-Qubit Gate Realization Using Single Quantum Particle}
\author{Alexander R. Kessel and Vladimir L. Ermakov \thanks{e-mail: ermakov@sci.kcn.ru}\\
{\it Kazan Physico-Technical Institute, }\\
{\it Kazan Science Center, }\\
{\it Russian Academy of Sciences,} \\
{\it Kazan 420029 Russia}}
\date{\today}
\maketitle
\begin{abstract}
Using virtual spin formalism it is shown that a quantum particle with eight
energy levels can store three qubits. The formalism allows to
realize a universal set of quantum gates. Feasible formalism implementation
is suggested which uses nuclear spin-7/2 as a storage medium and
radio frequency pulses as the gates. One pulse realization of all universal
gates has been found, including three-qubit Toffoli gate.
\end{abstract}


%\pacs{03.65.-w, 03.67.-a, 76.60.-k}
% 03.65.-w fundamental quantum mechanics
% 03.67.-a quantum information
% 76.60.-k Nuclear magnetic resonance and relaxation

\section{ Introduction}

\par Toffoli proved \cite{Toffoli} that a classical reversible computer can be constructed 
using a universal three-bit gate CCNOT (``controlled controlled NOT'' or ``double
controlled NOT''). Deutsch has shown that in quantum information theory the
generalized CCNOT gate - CCUT gate (``controlled controlled unitary
transformation'' or ``double controlled unitary transformation'') \cite{Deutsch} -
also is a universal one. However to realize CCUT is not an easy task since
there is no three body interaction in Nature. A way round was found Barenco {\it et al.} 
\cite{Barenco}: it was proved that CCUT can be realized using five two qubit CUT 
(``controlled unitary transformation'') gates. 
Such an approach was used in the NMR implementation of Toffoli gate
\cite[Eq. (38) and (39)]{CoryPriceHavel}.
\par The virtual spin formalism developed in \cite{KesselErmakov} is used in the present
paper to physically implement CCUT in a simplest possible way -
using one pulse and one quantum particle.

\section{The quantum information notation for gates of system of three real spins-$1/2$} 

\par In quantum information the gates for the three real spins-1/2 system usually are 
written in Hilbert space $\Gamma$ which is a direct product 
$\Gamma = \Gamma _{Q} \otimes \Gamma _{R}\otimes \Gamma _{S}$ 
of the Hilbert spaces 
of three real spins-1/2 $Q$, $R$, $S$. As a basis of $\Gamma$ the following eight 
functions can be chosen which will be used later 
\begin{equation} \label{GammaBasis}
\begin{array}{cccc}  
\ket{0}= \ket{000},&\ket{1}=\ket{001},& \ket{2}=\ket{010},& \ket{3}=\ket{011},\\
\ket{4}=\ket{100},& \ket{5}=\ket{101},& \ket{6}=\ket{110}, &\ket{7}=\ket{111},
\end{array}
\end{equation}
where $\ket{M}=\ket{m_{Q}, m_{R}, m_{S}}$, for example, \\
$\ket{5}=\ket{m_{Q}=+1/2, m_{R}= -1/2, m_{S}= +1/2}$ and so on.

\par Let us consider all possible gates with NOT operation. In three spin system there 
are three NOT gates - the gate, which invert one spin leaving two others unchanged, 

\begin{eqnarray} \label{NOTqrs}
NOT_{Q} &=& \ket{1 m_{R}m_{S}}\bra{0 m_{R} m_{S}} + \ket{0 m_{R} m_{S}}\bra{1 m_{R}m_{S}},\nonumber \\ 
NOT_{R} &=& \ket{m_{Q}1 m_{S}}\bra{m_{Q}0 m_{S}} + \ket{m_{Q}0 m_{S}}\bra{m_{Q}1 m_{S}},\\ 
NOT_{S} &=& \ket{m_{Q}m_{R} 1}\bra{m_{Q}m_{R}0} + \ket{m_{Q}m_{R}0}\bra{m_{Q}m_{R}1}.\nonumber 
\end{eqnarray}

\par There are six CNOT gates. For example, when ``master'' spin R state controls 
``slave'' spin Q state leaving spin S state unchanged

\begin{eqnarray}  \label{CNOTrq}
CNOT_{R\to Q}&=&\ket{00m_{S}}\bra{00m_{S}} + \ket{11m_{S}}\bra{01m_{S}} + \nonumber \\
&&\ket{10m_{S}}\bra{10m_{S}} + \ket{01m_{S}}\bra{11m_{S}}.
\end{eqnarray}

\noindent Reverse gate 
\begin{eqnarray}  \label{CNOTqr}
CNOT_{Q\to R}&=&\ket{00m_{S}}\bra{00m_{S}} + \ket{01m_{S}}\bra{01m_{S}}+ \nonumber \\ 
&&\ket{11m_{S}}\bra{10m_{S}} + \ket{10m_{S}}\bra{11m_{S}}.
\end{eqnarray}

\noindent Other four CNOT gates can be written for pairs $RS$ and $QS$ by analogy.

\par There are three CCNOT gates. For example, when two ``master'' spins $Q$ and $R$ 
control one ``slave'' $S$ one has 
\begin{equation}\label{CCNOTqrs}
\begin{array}{rcl}  
CCNOT_{Q,R\to S} &=& \ket{000}\bra{000} + \ket{001}\bra{001}+\\ 
&&\ket{010}\bra{010}+ \ket{011}\bra{011}+ \\
&&\ket{100}\bra{100}+ \ket{101}\bra{101}+ \\
&&\ket{111}\bra{110}+ \ket{110}\bra{111}.
\end{array}
\end{equation}

\noindent The gates $CCNOT_{R,S\to Q}$ and $CCNOT_{S,Q\to R}$ can be written by analogy. 

\section{Virtual spin formalism}

In the realized NMR quantum gates one qubit NOT operation is implemented as 
a spin rotation under influence of a resonant pulse. 
Whereas conditional quantum 
dynamics, which is necessary for two qubit CNOT gate, is implemented using 
system with two body spin-spin interaction \cite{CoryPriceHavel}. 
\par It will be shown later in this paper that conditional quantum dynamics can be 
realized on one spin-7/2 without using spin-spin interactions. 
Another specific feature of such an approach is 
that all three types of gates NOT, CNOT and CCNOT can be implemented using just one RF pulse. These advantages are 
possible due to special information coding onto spin-7/2 states - this coding was 
introduced in \cite{KesselErmakov} under the name virtual spin formalism. 
\par The main idea of the virtual spin formalism can be expressed as follows. As a 
basis for the spin-7/2 Hilbert space $\Gamma _{7/2}$ the eigen functions 
$\ket{\chi _{m}}$ of $I_{z}$ 
operator 
$(m= \pm \frac{1}{2},\pm \frac{3}{2}, \pm \frac{5}{2}, \pm \frac{7}{2})$ 
or eigen functions 
$\ket{\psi _{m}}$ of total spin energy operator (defined below) 
can be chosen. Let us instead 
$$m = -\frac{7}{2}, -\frac{5}{2}, -\frac{3}{2}, -\frac{1}{2}, 
+\frac{1}{2}, +\frac{3}{2}, +\frac{5}{2}, +\frac{7}{2}$$ 
use notation 
$M=0,1, 2, 3, 4, 5, 6, 7$  
and therefore instead $\ket{\psi _{m}}$ to use $\ket{\psi _{M}}$. 
Then formally $\ket{\psi _{M}}$ can be mapped to $\ket{M}=\ket{m_{Q}, m_{R}, m_{S}}$ which is a 
state of virtual spins-1/2 $Q$, $R$, $S$. Then in order to implement all above 
mentioned gates it is necessary to find such external influence on real spin-7/2, which 
would have in the basis $\ket{\psi _{M}}$ the evolution propagator matrices of the same 
operational forms as the above written gates in the basis $\ket{M}$. 

\section{Spin-7/2 system and physical realization of gates}

Let us consider the system of nuclear spin $I=7/2$ placed in constant magnetic 
field $H_{0}$ and axially symmetrical electric crystal field:

\begin{equation}\label{Hamiltonian}
\begin{array}{ll}
{\cal H} = {\cal H}_{z} + {\cal H}_{Q},& \\
{\cal H}_{z} = - \hbar \omega _{0} I_{z},& 	
{\cal H}_{Q} = \frac{1}{2} \hbar \omega _{Q} \sum \limits _{a=0, \pm 1, \pm 2} Q_{\alpha} q_{-\alpha}, \\
Q_{0}=I_{z}^{2} - \frac{1}{3} I(I+1),&
q_{0}=3 \cos ^{2} \theta - 1,\\ 	

Q_{\pm 1}=I_{z}I_{\pm 1} + I_{\pm 1}I_{z},&
q_{\pm 1}=\sin \theta \cos \theta e^{\pm i\phi},\\
Q_{\pm 2}=I_{\pm 1}^{2},&
q_{\pm 2}= \frac{1}{2} \sin 2 \theta e^{\pm i 2 \phi},\\
\omega _{Q}= 
\frac
{\displaystyle 3 e q Q}
{\displaystyle [2I(2I-1) \hbar]}, &
I_{\pm 1} = I_{x} \pm i I_{y}, 
\end{array}
\end{equation}

\noindent where $e$ - the electron charge, $Q$ - nuclear quadrupole moment, 
$I_{\beta} (\beta=x,y,z)$ - spin 
components, $2eq$ - the electric field gradient value, 
$\theta$ and $\phi$ - the polar angles, which 
define its axes orientation in laboratory system frame. 
Let us consider a case when $\omega _{Q} << \omega _{0}$ 
so quadrupole interaction influence can be calculated using perturbation theory. It is 
supposed that quadrupole interaction much greater than the spin resonance line width, 
so the spectrum consists of seven well separated resonance lines.

The first approximation gives the following energy levels and eigen functions:
\begin{equation}\label{EigenValFun}
\begin{array}{l}
E_{m} \equiv \hbar \varepsilon _{m} = 
- \hbar \omega _{0}m + \hbar \omega _{Q} q_{0} (m^{2} - \frac{21}{4}),\\
\ket{\psi _{m}} = \chi _{m} + 
\sum \limits _{m \neq k} \frac 
{\displaystyle \bra {\chi _{k}} {\cal H}_{Q} \ket{\chi _{m}}}
{\displaystyle \hbar \omega _{0}(k - m)} 
\chi _{k} , 
\end{array}
\end{equation}

\noindent here the normalization factor of $\ket{\psi _{m}}$ is omitted. 

For simplicity the projective operators $P_{mn}$ notation will be used. 
They are matrices 
$8 \times 8$ with all elements $p_{kl}$ equal zero except one $p_{mn}=1$. 
They have very simple relations:

\begin{equation}  \label{Projective}
P_{kl}P_{mn} = \delta _{lm}P_{kn} ,	
P_{mn} = P_{nm}^{+} , 	
P_{mn}\ket{\psi _{k}} = \delta _{nk}\ket{\psi _{m}}. 
\end{equation}

\par An RF pulse, which is resonant for energy levels $E_{m}$ and $E_{n} (E_{m} > E_{n})$, 
results in evolution operator \cite{KesselErmakov}

\begin{equation}  \label{Vxmnf}
\begin{array}{l}
V_{X}(\phi_{mn}, f  ) = \hat 1 +(P_{nn} + P_{mm})(\cos \frac{1}{2} \phi _{mn} - 1 ) +\\ 
\qquad i(P_{mn} e^{if} + P_{nm} e^{-if}) \sin \frac{1}{2} \phi _{mn},\\
\phi _{mn} = 2(t-t_{0}) \gamma H_{rf} \vert \bra{n}I_{x} \ket{m} \vert,
\hat 1 = \sum _{m} P_{mm}, 
\end{array}
\end{equation}

\noindent where oscillating magnetic field is parallel to $X$ axis, and $H_{rf}$, $f$ and 
$\Omega (=\Omega _{mn}\equiv (E_{m} -E_{n})/\hbar)$ - its amplitude, phase and frequency, 
and $\hat 1$ - unit operator in space $\Gamma _{7/2}$. A case 
when the field is parallel to $Y$ axis results in replacing $f$ by $f + \frac{1}{2}\pi$ 
in (\ref{Vxmnf}). 

\par Let us consider the realization of logic gates on the spin-7/2 states in order of 
increasing their complexity. 

\par The CCNOT gate requires one single frequency pulse. For example, the gate 
$CCNOT_{Q,R\to S}$ is realized using pulse with frequency 
$\Omega _{67}$ and with rotation angle $\pi$. 
According to (\ref{Vxmnf}) the 
evolution operator at such conditions has the following form
\begin{equation}  \label{Vxpi670}
V_{X}(\pi _{67}, 0) = \hat 1 - (P_{77} + P_{66})+ i(P_{67} + P_{76}).
\end{equation}

\noindent Taking into account the above mentioned isomorphism one can see that the following 
equality takes place: 
\begin{equation}  \label{P6776}
P_{67} + P_{76} = \ket{6}\bra{7} + \ket{7}\bra{6} = \ket{110}\bra{111} + \ket{111}\bra{110}.
\end{equation}

\noindent It means that the matrix of evolution operator $V_{X}(\pi _{67}, 0)$ 
is equal to matrix of the gate 
$CCNOT_{Q,R\to S}$ (Eq. (\ref{CCNOTqrs})):

\begin{equation}  \label{VxCCNOT}
V_{X}(\pi _{67}, 0) = CCNOT_{Q,R\to S}
\end{equation}

\noindent up to the phase factor $i$ for non diagonal elements. For other two gates one has
\begin{equation}  \label{Vx75}
\begin{array}{l}
V_{X}(\pi _{75}, 0) = CCNOT_{Q,S\to R},\\ 		
V_{X}(\pi _{73}, 0) = CCNOT_{R,S \to Q}.
\end{array}
\end{equation}

It is necessary to point out that in comparison with transition $\Omega _{67}$, 
the transitions $\Omega _{57}$ and $\Omega _{37}$ between the states $\chi _{M}$ 
are forbidden and become non zero in the first order of 
parameter $\omega _{Q}/\omega _{0}$. To get the $\pi$ rotation in these 
cases longer pulses or stronger RF field 
are necessary. However, numerical calculations show, that when 
$\omega _{Q}$ and $\omega _{0}$ are of the 
same order of magnitude, the expressions for rotation angles also have the matrix 
elements of the same orders. 
\par The CNOT gate requires one double frequency pulse, the evolution operator of 
which is a product of two operators:
\begin{equation}  \label{VxVx}
\begin{array}{l}
V_{X}(\pi _{23}, 0) V_{X}(\pi _{67}, 0) = CNOT_{R\to S},\\
V_{X}(\pi _{13}, 0) V_{X}(\pi _{57}, 0) = CNOT_{S\to R},\\
V_{X}(\pi _{45}, 0) V_{X}(\pi _{67}, 0) = CNOT_{Q\to S},\\
V_{X}(\pi _{15}, 0) V_{X}(\pi _{37}, 0) = CNOT_{S\to Q},\\
V_{X}(\pi _{46}, 0) V_{X}(\pi _{57}, 0) = CNOT_{Q\to R},\\
V_{X}(\pi _{26}, 0) V_{X}(\pi _{37}, 0) = CNOT_{R\to Q}.
\end{array}
\end{equation}

The NOT gate requires one four-frequency pulse, the evolution operator of which is a 
product of four operators:
\begin{equation}  \label{VxVxVxVx}
\begin{array}{l}
V_{X}(\pi _{04}, 0) V_{X}(\pi _{15}, 0) V_{X}(\pi _{26}, 0) V_{X}(\pi _{37}, 0)= NOT_{Q},\\
V_{X}(\pi _{02}, 0) V_{X}(\pi _{13}, 0) V_{X}(\pi _{46}, 0) V_{X}(\pi _{57}, 0)= NOT_{R},\\
V_{X}(\pi _{01}, 0) V_{X}(\pi _{23}, 0) V_{X}(\pi _{45}, 0) V_{X}(\pi _{67}, 0)= NOT_{S}.
\end{array}
\end{equation}

The physical realizations of gates, expressed in Eq. (\ref{VxCCNOT})-(\ref{VxVxVxVx}), in 
comparison with adopted in quantum information notation have additional phase factor $i$ 
for non diagonal elements. It should be taken into account later, when these gates will be 
used for constructing complex algorithms. 

\par Above for simplicity specific values of parameters $\phi$ and $f$ have been used in 
evolution operators. The expressions (\ref{VxCCNOT})-(\ref{VxVxVxVx}) for CCNOT, CNOT, NOT 
can be easily 
generalized to get expressions for CCUT, CUT, UT. For example, if one uses Eq. (\ref{Vxmnf}) 
with arbitrary parameters $\phi$ and $f$, then (\ref{VxCCNOT}) gives the expression for CCUT. 
The corresponding calculations are straightforward but rather complicated, and would hide 
the main idea of the paper.


\begin{thebibliography}{9}

\bibitem{Toffoli}  
T. Toffoli, 
{\it ``Reversible Computing''},
In: Automata, Languages and Programming
(editors: J. W. de Bakker and J. van Leeuwen, Springer, New York, 1980),
pp. 632-644.

\bibitem{Deutsch}  
D. Deutsch, 
{\it ``Quantum computational networks''},
Proc. Roy. Soc. Lond. {\bf A425}, 73-90 (1989).

\bibitem{Barenco}  
Adriano Barenco,
Charles H. Bennett,
Richard Cleve,
David P. DiVincenzo,
Norman Margolus,
Peter Shor,
Tycho Sleator,
John Smolin,
Harald Weinfurter,
{\it ``Elementary gates for quantum computation''},
Phys. Rev. {\bf A52}, 3457 (1995), 
quant-ph/9503016.

\bibitem{CoryPriceHavel}  
David G. Cory, 
Mark D. Price,
Timothy F. Havel,
{\it ``Nuclear Magnetic Resonance Spectroscopy:
An Experimentally Accessible Paradigm for
Quantum Computing''},
Physica D, {\bf 120}, 82 (1998), 
quant-ph/9709001.


\bibitem{KesselErmakov} 
Alexander R. Kessel, 
Vladimir L. Ermakov,
{\it ``Multiqubit Spin''}, 
JETP Letters, Vol. {\bf 70},  pp. 61-65 (1999),
(translation from Russian: 
Pis'ma v Zh. Eksp. Teor. Fiz. Vol. {\bf 70}, No. 1, pp. 59-63, 
10 July 1999), 
quant-ph/9912047.

\end{thebibliography}
\end{document}